\documentclass[a4paper]{jpconf} 
\usepackage{graphicx}
\usepackage{caption}
\usepackage{subcaption}
\usepackage{hyperref}
\usepackage{float}
\begin{document}
\title{\centering{Electronic and Band Structure calculation of Wurtzite CdS Using GGA and GGA+U functionals}}

\author{Ankan Biswas$^1$, S. R. Meher$^2$, Deepak K. Kaushik$^{1*}$}

\address{$^1$Department of Physics, Lovely Professional University, Phagwara 144001, India.\\

$^2$Department of Physics, School of Advanced Sciences, VIT University, Vellore 632014, India.}

\ead{$^*$kaushikd189@gmail.com}

\begin{abstract}
The wurtzite (wz) structure of CdS is analysed using density functional theory within the generalized gradient approximation (GGA) and Hubbard correction (GGA+U). The total energy convergence evaluation is carried out concerning energy cut-off (ecutwfc) and k-point sampling. The geometry optimization of wz-CdS is calculated using the total energy and force minimization process, which is based on the Broyden–Fletcher–Goldfarb–Shanno (BFGS) optimization algorithm. Bulk modulus and lattice parameters are estimated to ensure accuracy of the calculations. The electronic band structure, density of states (DOS), and projected density of states (PDOS) of wz-CdS are analysed. The band structure calculation shows CdS as direct band gap semiconductor. The electronic correlation in CdS is altered by varying U-parameters of valence orbitals of Cd and S. The alteration of electronic correlation results in convergence of the band gap to the experimental value 2.4 eV. The alteration of U-parameter affects substantially the density of states near the band edges.
\end{abstract}

\section{Introduction}

\vspace*{0.3cm}
Cadmium sulfide (CdS), a $II-VI$ group semiconductor, has potential applications in optoelectronic devices \cite{priyam2007surface}. Due to direct band gap of 2.42 eV (bulk CdS) \cite{liu2016lattice} and size dependent optical response, CdS is used in photodetectors \cite{li2015hierarchical}, photocatalysis \cite{wang2020two}, quantum-dot LEDs \cite{zhang2017significant}, thin-film transistors (TFTs) \cite{voss2004cadmium} etc. Being n-type semiconductor (due to sulfur vacancy) \cite{meher2017native}, CdS is considered as a suitable buffer layer with $CdTe$, $Cu(In, Ga)Se_2$ and $Cu_2ZnSn(S,Se)_4$ absorbers for thin film solar cells\cite{hur2008enhancement, cheng2020efficiency}. Moreover, CdS offers minimum lattice mismatch with these absorbers and exhibits low minority carrier recombination at the absorber/buffer interface. Due to thermal and chemical stability, it is also used in color pigments \cite{van2012combined}. CdS crystallizes in three phases; zinc blende (zb), wurtzite (wz) and rock salt (rs) which strongly depend on pressure and temperature \cite{xiong2008growth, guler2017elastic}. Among these phases, wz-CdS is widely applicable for optoelectonic devices.

\vspace*{0.3cm}
Density functional theory (DFT) is a popular ab-initio approach for calculating electronic properties of a crystal structure \cite{burke2012perspective}. The Kohn–Sham DFT formalism along with local density approximation (LDA) or generalized gradient approximation (GGA), has certain limitations. Electronic correlations and exchange interactions are not effectively accounted in such approximations. On the other hand, Hartree–Fock (HF)-based approaches such as coupled cluster \cite{oliphant1994systematic} or higher-order perturbation theory \cite{pople1976theoretical}, provide superior findings but are confined to the smaller molecules. A common extension of DFT called DFT + U, where U is a Hubbard potential added to the Kohn–Sham Hamiltonian, is a significantly fast technique \cite{finazzi2008excess}. The DFT+U approach is anticipated to be a suitable framework for analyzing crystals containing 3d-transition metals and, in some cases, 5f actinides metals \cite{bouchet2000equilibrium}.

\section{Computational Method}

\vspace*{0.3cm}
In the present article, the structural and electronic properties of wz-CdS were investigated by density functional theory (DFT) calculations using projector augmented plane wave (PAW) method \cite{gajdovs2006linear} applying semilocal Perdew-Burke-Ernzerhof (PBE) exchange correlation functional \cite{perdew1996generalized}.The electron wave functions were represented in plane wave basis truncated at kinetic energy of 60 Ry and the charge density cutoff of 600 Ry. The calculations were employed in the Quantum ESPRESSO (version 6.5) which is an open source suite of codes\cite{giannozzi2017advanced,giannozzi2009quantum}. Graphical user interfaces such as BURAI and XCrySDen were used to visualize crystal structure. A mesh of $10\times10\times6$ k-points was employed for the Brillouin zone integration \cite{monkhorst1976special}. Occupation with Marzari-Vanderbilt smearing \cite{marzari1999thermal} was used for self consistent field calculation. Similarly,  tetrahedra occupation \cite{robouch2002statistical} was used for the non self consistent filed calculation with smearing parameter of 0.001 Ry. Geometry optimization was performed using the BFGS algorithm by setting threshold energy less than $10^{-6}$ Ry and force less than $10^{-4} $ Ry/Bohr \cite{yuan1991modified}. The electronic band gap estimated by PBE functional was found to be less than the experimental values. In order to improve the results, Hubbard-U parameter was introduced in Cd-d and S-p orbitals for DFT+U calculations.

\section{Result and Discussion}

\subsection{Crystal structure}

\vspace*{0.3cm}

\begin{figure}[H]
\centering
\begin{minipage}{14pc}
\includegraphics[scale=0.25]{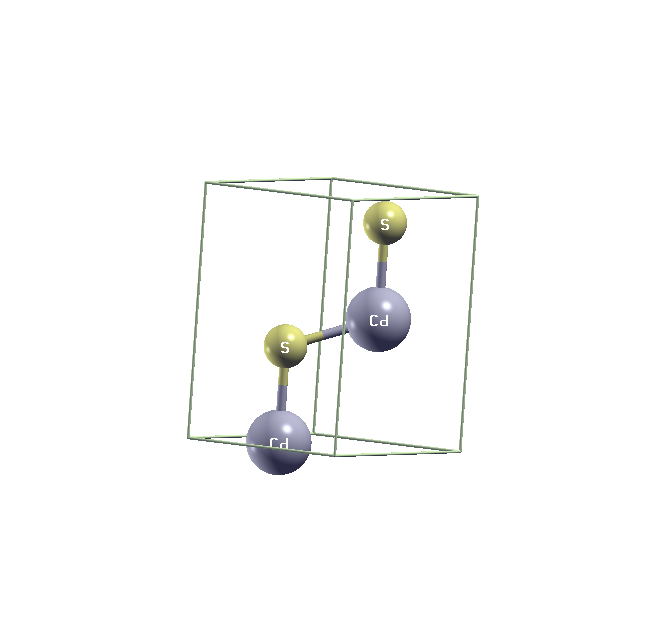}
\subcaption{}
\label{fig:x bvc}
\end{minipage}\hspace{1pc}
\begin{minipage}{14pc}
\includegraphics[scale=0.25]{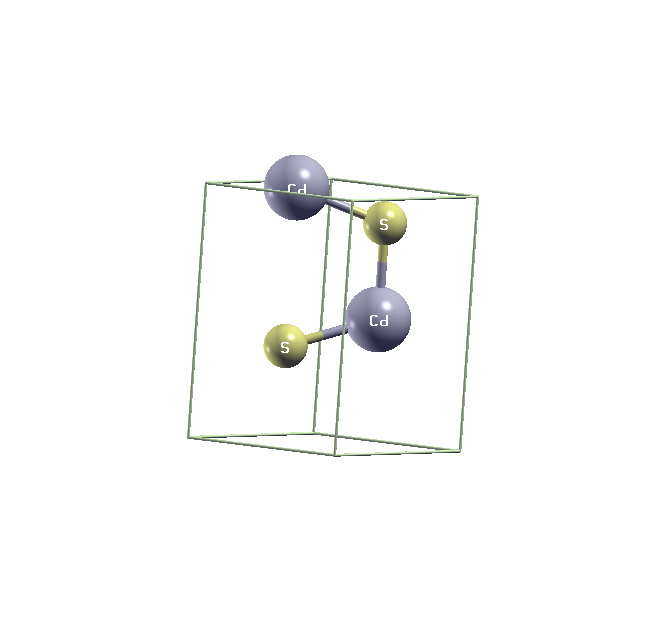}
\subcaption{}
\label{fig:x avc}
\end{minipage}
\caption{\label{fig:x crystal}Crystal structure of CdS (a) before relaxation and (b) after relaxation.}

\end{figure}

Fig. \ref{fig:x crystal} depicts the optimized crystal structure of wz-CdS (space group $P6_3mc$) as visualized by XCrySDen software. The unit cell consists of two Cd and two S atoms. To obtain optimal lattice, atomic positions and volume of CdS is relaxed. Fig. \ref{fig:x bvc} and Fig. \ref{fig:x avc} correspond to the crystal structures before and after relaxation respectively. The bond length of Cd-S changes from $2.57\AA$ to $2.55\AA$ after relaxation which is similar to the reported experimental value \cite{heiba2019effect}. Table ~\ref{lattice} shows that the lattice parameters obtained from DFT+U calculations are converged to the experimental results \cite{zhou2014revisiting}. The calculated bulk modulus is 61.7 GPa, which is also similar to the experimental results \cite{wright2004interatomic}.

\begin{center}
\begin{table}
\caption{\label{lattice}Lattice parameters of wz-CdS obtained from GGA and GGA+U calculations.}
\footnotesize
\centering
\begin{tabular}{@{}l*{7}{l}}
\br
&Lattice parameters    &Experimental (\AA)\cite{zhou2014revisiting} &        GGA (\AA)       & Error* (\%)	&      GGA+U (\AA)	& Error (\%)  \\
\mr
&a &4.136	&4.201	& 1.57	&4.171	& 0.85\\

&c &6.714	&6.828  & 1.70	&6.710 	& 0.06\\
\br
\end{tabular}\\
$^*$ The parameters are compared with the experimental values.
\end{table}
\end{center}
\normalsize

\subsection{Band-structure}

\begin{figure}[H]
\centering
\begin{minipage}{18pc}
\includegraphics[scale=0.25]{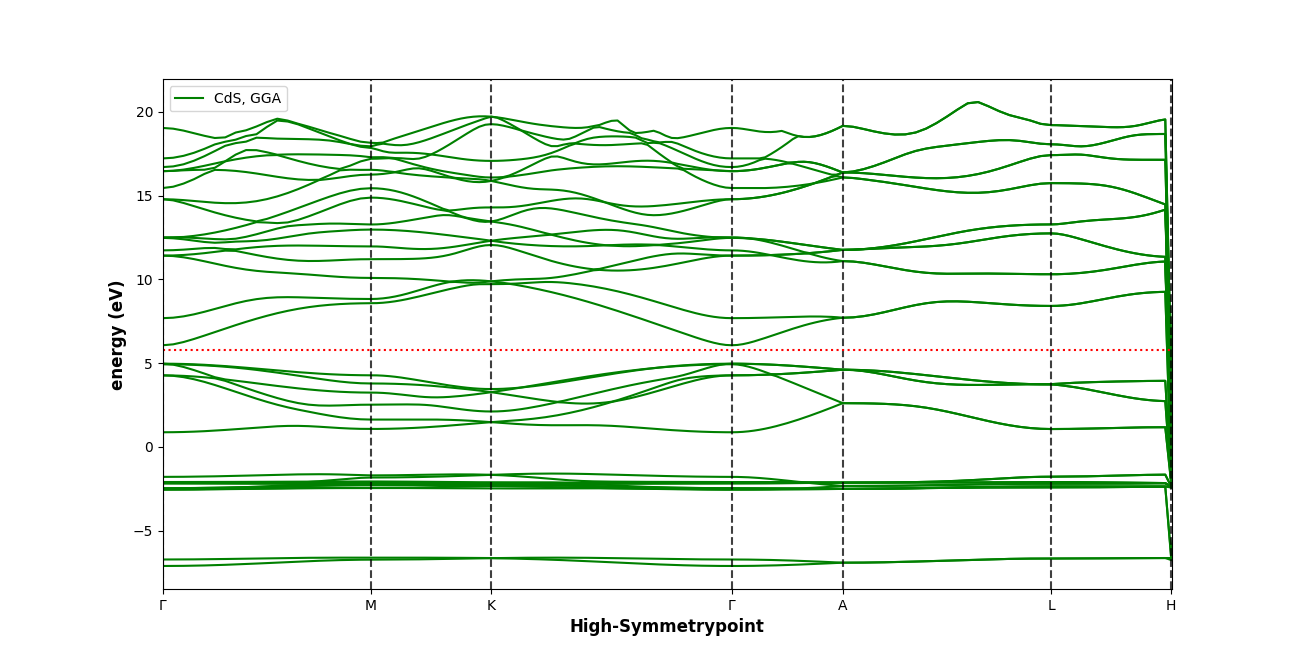}
\subcaption{}
\label{fig:x bands}
\end{minipage}\hspace{4.6pc}
\begin{minipage}{14pc}
\includegraphics[scale=0.23]{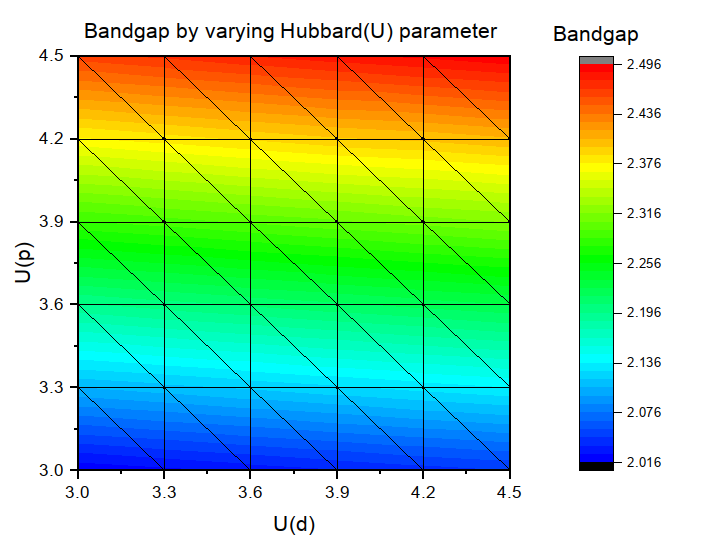}
\subcaption{}
\label{fig:x Contour}
\end{minipage}
\begin{minipage}{19pc}
\includegraphics[scale=0.25]{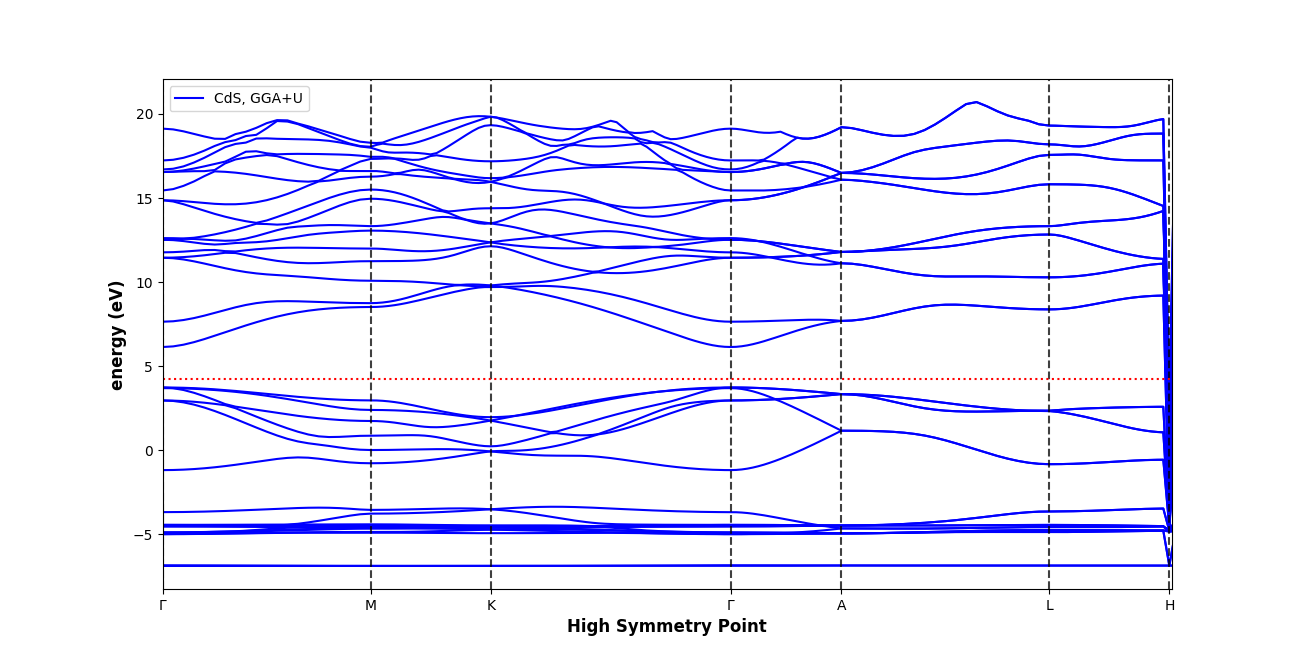}
\subcaption{}
\label{fig:x bands-u}
\end{minipage}
\begin{minipage}{18.5pc}
\includegraphics[scale=0.25]{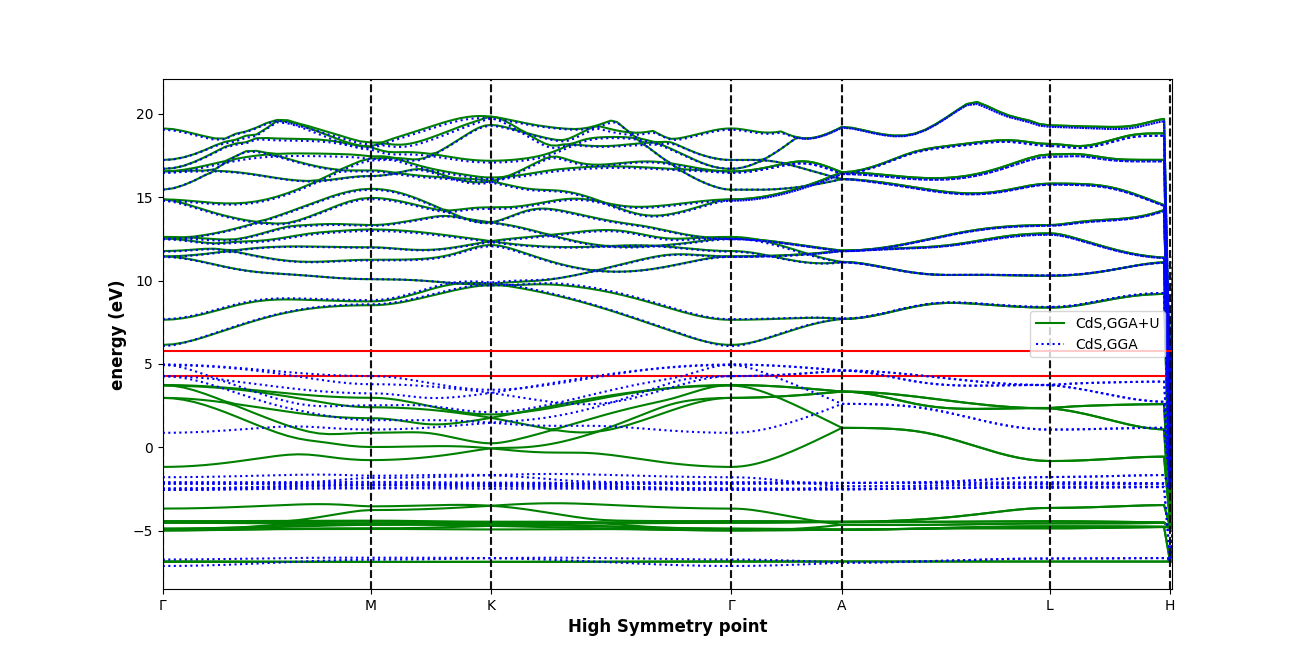}
\subcaption{}
\label{fig:x bands-both}
\end{minipage}\hspace{4pc}
\caption{(a) Band structure plot of wz-CdS calculated using GGA functional (b) contour plot of bandgap at different $U_d$ and $U_p$ parameters (c) band structure plot calculated using GGA+U functional for $U_d = 4.5$ and $U_p = 4.2$ (d) comparison of band structures calculated using GGA (dotted lines) and GGA+U (solid lines) functionals}
\end{figure}

\vspace*{0.3cm}
Band structure calculation of a semiconductor is important to estimate band gap, nature of energy bands and their dispersion. These properties are advantageous for optoelectronic devices such as LEDs, solar cells, photodetectors etc. \cite{fahrenbruch2012fundamentals}. Fig. \ref{fig:x bands} shows the band structure plot of wz-CdS calculated using PBE-GGA functional. The bandgap, extracted from the the difference of valance-band maxima and conduction band minima at high symmetry $\Gamma$ point, is 1.10 ev which is $56\%$ lesser than the experimental value\cite{madelung2004semiconductors}. The under-estimation of band gap is a common error in the DFT calculations \cite{cheng2016first}. Traditional DFT calculations fail to account for some electronic 'correlations' in d-orbitals, such as self-interaction corrections (SIC). Any orbital that receives a positive U (Hubbard parameter) is further localized by increasing intra-band repulsive Coulomb interaction. In order to improve the band gap calculations, the DFT+U approach \cite{aras2014combined} is introduced by varying $U_d$ as well as $U_p$ for Cd-d and S-p orbitals respectively. The contour plot as shown in Fig. \ref{fig:x Contour} illustrates that $U_p$ plays a crucial role in improving the bandgap value. For $U_p \geq 4.2 $, bandgap is more than 2.3 eV for any $U_d \in [3.0, 4.5]$.

Fig. \ref{fig:x bands-u} shows the band structure of wz-CdS using DFT+U calculations for $U_d = 4.5$ and $U_p = 4.2$. The bandgap obtained is 2.40eV, which is well converged to the experimental values \cite{madelung1982numerical}. The substantial change in the bandgap is due to downward shift in valance band maxima (VBM) and the valance band as well (Fig. \ref{fig:x bands-both}). There is no significant change observed in the conduction band. The valance band maxima at the $\Gamma$-point of a CdS unit cell are made up of two strongly scattering degenerate bands. Adding U contribution to the orbitals, the Fermi energy shifted downward $(5.78 eV-4.27 eV)$ i.e 1.51 eV. The heavy overlapping of S-p orbitals with Cd-p orbitals causes shifting of valance band maxima.

The upper valance band and lower valance band width almost 5 eV and 2 eV. The S-p orbital dominates in the top valance band portion, whereas the Cd-d orbital dominates in the lower valance band. Conduction band minima play an important role in demonstrating semiconductor conductivity and optical properties. At the $\Gamma$-point, the conduction band consists of a single scatter band. The width of this band is nearly 3 eV. Conduction band minima are greatly influenced by the Cd-s and S-p orbitals.The top conduction band is dominated by the Cd-p orbital, whereas the bottom conduction band is dominated by the S-p orbital.

\subsection{Density of state and Projected Density of state}

\begin{figure}[H]
\centering
\begin{minipage}{27pc}
\includegraphics[scale=0.30]{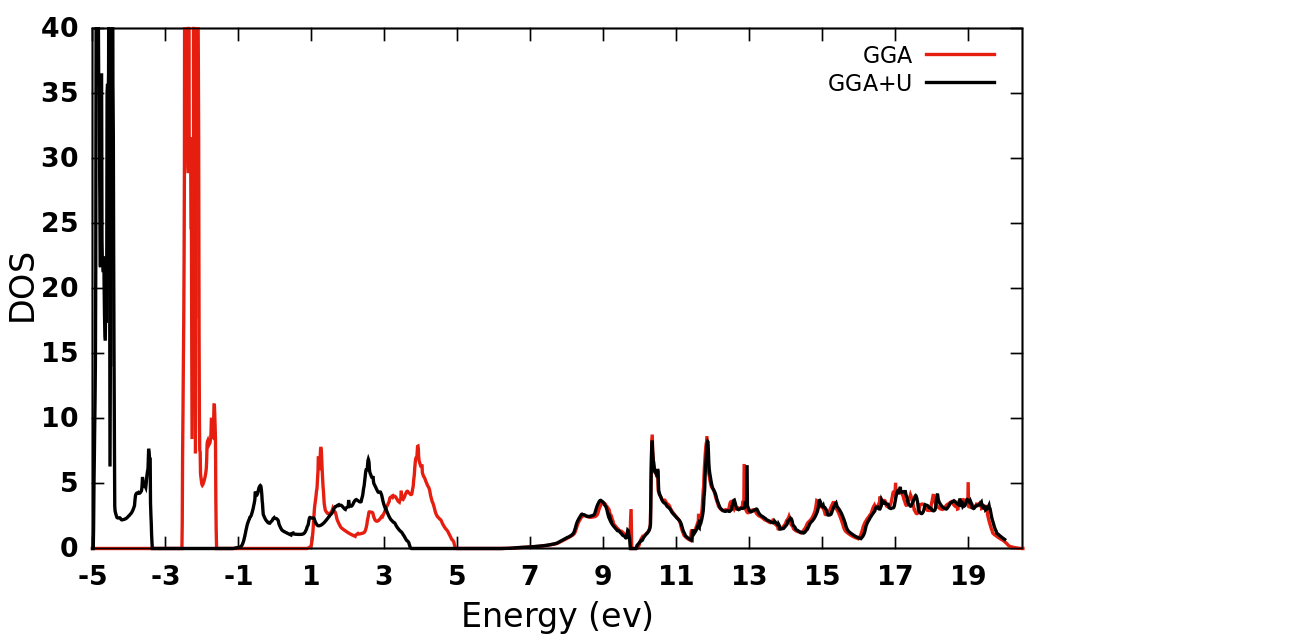}
\caption{\centering{Density of state with GGA method and GGA+U method.}}
\label{fig:x dos}
\end{minipage}
\end{figure}

\begin{figure}[H]
\begin{minipage}{18pc}
\includegraphics[scale=0.21]{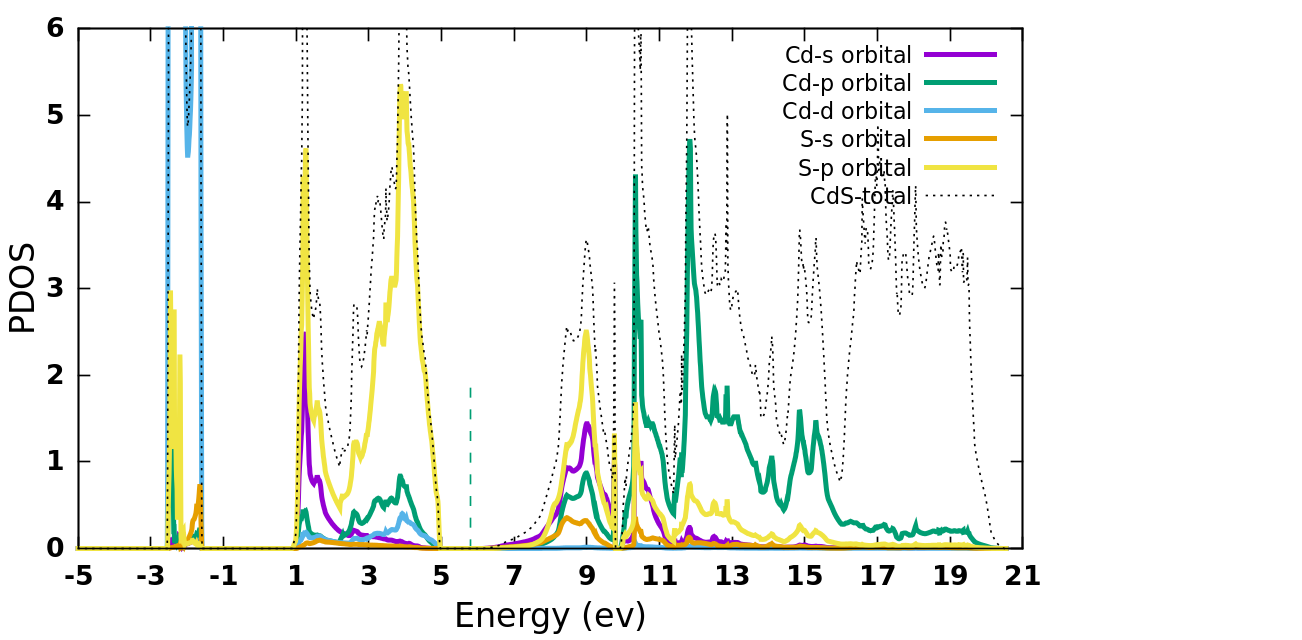}
\subcaption{\centering{}}
\label{fig:x dos1}
\end{minipage}\hspace{1pc}
\begin{minipage}{19pc}
\includegraphics[scale=0.21]{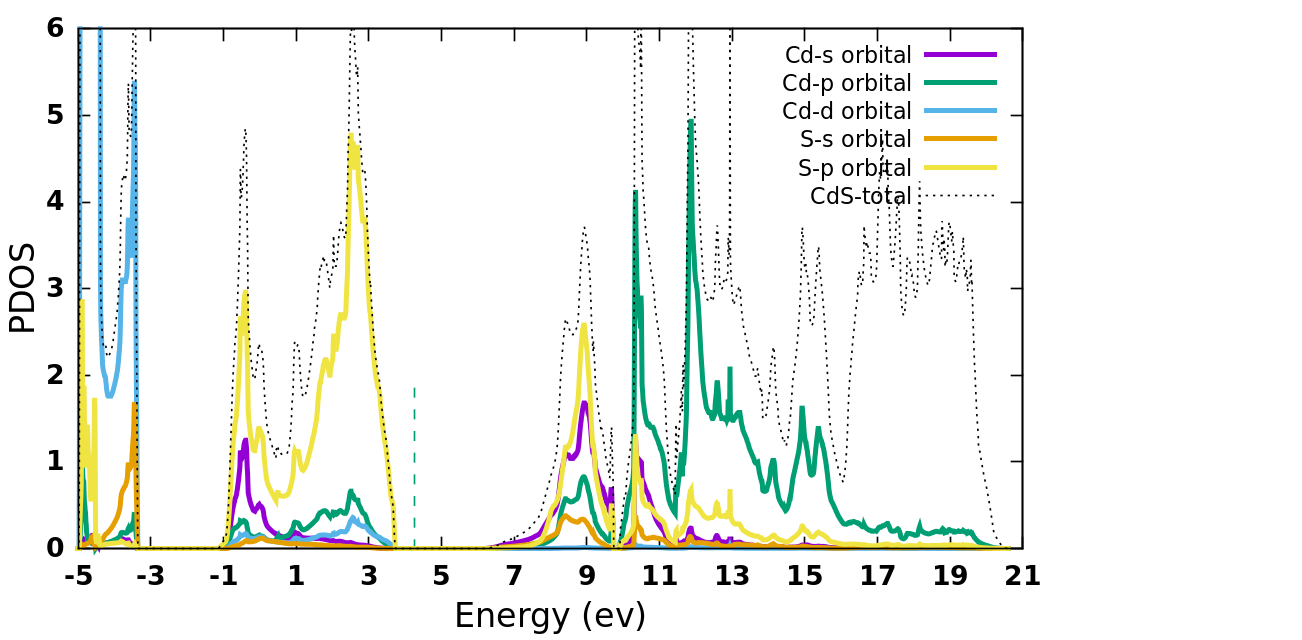}
\subcaption{\centering{}}
\label{fig:x dos-u}
\end{minipage}
\caption{Projected density of state contribution of all orbitals with (a) GGA method and (b) GGA+U method.}
\end{figure}

\vspace*{0.3cm}
The density of states (DOS) determines the number of accessible energy states or levels per unit volume for electrons in the conduction band and holes in the valance band. In general, p-orbitals are delocalized in nature, with less intra-band interaction between electrons in 3d orbitals\cite {khatri2014application}.
Fig. \ref{fig:x dos} show DOS calculated using GGA and GGA+U methods. The valance band is mostly composed of S-p, Cd-p, Cd-d, and Cd-s orbitals. The valance band has a significant contribution due to S-p orbital and is peaked at around 4 eV  as shown in Fig. \ref{fig:x dos-u}. S-s and Cd-d orbitals have minimal contributions in the valance region. In CdS, S-p orbital is partially filled and contributes significantly to the valence band maximum.  By introducing additional Coulomb repulsive interaction to S-p orbital with U values, the orbital becomes more localized. It is vital to investigate whether the U would play a significant role in the S-p orbital. If the orbital is deep in the valence band, such as the 4d orbital in Cd, the addition of a minor correlation via U does not affect the electronic properties of CdS (Fig. \ref{fig:x dos1}). On the other hand, p-bands dominate the top of the valence band for CdS. By including the U parameter, p-orbital is pushed towards the lower energy side.
The conduction band of wz-CdS is dominated by Cd-p orbitals, along with small contributions from Cd-d,s, and S-s orbitals. It is observed from the Fig. \ref{fig:x dos} that the valance band edge is shifted to the lower energy by 1 eV approximately, but the conduction band edge is not altered.  The DOS calculated by both the techniques is compared in Fig. \ref{fig:x dos}. Inclusion of the U parameter leads to the shifting of valance band to the lower energy. The top and lower valance bands moved almost 1 eV apart. The bandgap of the material is obtained from the DOS plot where it becomes discontinuous between the top of the valance band and the bottom of the conduction band.


\section{Conclusion}

In summary, structural and electronic properties of wurtzite-CdS are studied using density functional theory and plane wave augmented method. The exchange-correlation functional is approximated using GGA and GGA+U. The crystal structure studies reveal that the lattice constant, bond length and bulk modulus obtained by GGA+U calculation are in agreement with the experimental results. The bandgap of wz-CdS is underestimated by GGA calculation. Further, the calculations are improved by introducing Hubbard parameter (U) in the GGA functional. The U-parameter is varied for Cd-d orbital and S-p orbitals. The optimal direct band gap around 2.40 eV is obtained at $U_d = 4.5$ and $U_p = 4.2$. It is evident from the band structure results that the valance band is shifted down by introducing U parameters. Moreover, the interaction U-parameter on S-p orbital influence substantially the density of states and hence the band structure. These results will be further implemented to understand electronic properties of un-doped and doped-CdS supercells.

\pagebreak



\section*{References}

\bibliography{iop}
\bibliographystyle{iop}

\end{document}